# Proactive DDoS Detection and Mitigation in Decentralized Software-Defined Networking via Port-Level Monitoring and Zero-Training Large Language Models

Mohammed N. Swileh (mohammedswileh2023@email.szu.edu.cn)[a,1], Shengli Zhang (zsl@szu.edu.cn)[a,*]

[a] *College of Electronics and Information Engineering, Shenzhen University, Shenzhen, 518060, China*

[1] *First author*

[*] *Corresponding author*

**Abstract**

Centralized Software-Defined Networking (cSDN) offers flexible and programmable control of networks but suffers from scalability and reliability issues due to its reliance on centralized controllers. Decentralized SDN (dSDN) alleviates these concerns by distributing control across multiple local controllers, yet this architecture remains highly vulnerable to Distributed Denial-of-Service (DDoS) attacks. In this paper, we propose a novel detection and mitigation framework tailored for dSDN environments. The framework leverages lightweight port-level statistics combined with prompt engineering and in-context learning, enabling the DeepSeek-v3 Large Language Model (LLM) to classify traffic as benign or malicious without requiring fine-tuning or retraining. Once an anomaly is detected, mitigation is enforced directly at the attacker's port, ensuring that malicious traffic is blocked at their origin while normal traffic remains unaffected. An automatic recovery mechanism restores normal operation after the attack inactivity, ensuring both security and availability. Experimental evaluation under diverse DDoS attack scenarios demonstrates that the proposed approach achieves near-perfect detection, with 99.99% accuracy, 99.97% precision, 100% recall, 99.98% F1-score, and an AUC of 1.0. These results highlight the effectiveness of combining distributed monitoring with zero-training LLM inference, providing a proactive and scalable defense mechanism for securing dSDN infrastructures against DDoS threats.

## 1. Introduction

Traditional computer networks have long suffered from rigidity and limited adaptability, as control and data planes are tightly coupled within proprietary hardware [1]. This tight integration restricts programmability, hinders innovation, and complicates network management, especially as traffic demands and cyber threats continue to grow [2, 3]. To address these challenges, the concept of Software-Defined Networking (SDN) emerged as a transformative paradigm [4, 5]. By decoupling the control plane from the data plane, SDN introduces a logically centralized controller that manages the entire network through programmable interfaces [6]. This separation not only simplifies network configuration and policy enforcement but also provides greater flexibility, scalability, and visibility compared to traditional architectures [7]. As a result, SDN has become a foundational technology and is now deployed by major companies such as Google, Facebook, Microsoft, and Amazon to manage their large-scale, dynamic networks [8].

Although cSDN simplifies network management by decoupling the control plane from the data plane, this very separation introduces new challenges [9]. The dependence on a single logically centralized controller creates a single point of failure, where any disruption to the controller can paralyze the entire network [10]. Furthermore, the detachment of the control plane from forwarding devices increases system complexity, leading to vulnerabilities such as availability issues, instability during failures, and performance degradation under heavy load [11, 12]. As the network grows in scale, the centralized controller also encounters scalability bottlenecks, struggling to process massive volumes of flow requests in real time. These inherent limitations raise significant concerns regarding the robustness and resilience of cSDN, especially when facing distributed and large-scale attacks [13].

To address the inherent limitations of cSDN, the concept of dSDN has recently emerged as a promising paradigm [14]. In dSDN, the control plane is no longer concentrated in a single controller but is distributed across multiple controllers, each operating closer to the data plane elements. This architectural shift mitigates the risk of a single point of failure, enhances network reliability and availability, and allows the system to continue functioning even if one controller fails [14]. Importantly, dSDN retains all the core benefits of cSDN (such as programmability, centralized policy enforcement, and global network visibility) while overcoming its drawbacks [14]. Furthermore, distributing the control logic enables better scalability, as the workload is shared among controllers, and improves performance by reducing latency through localized decision-making. As a result, dSDN represents a significant step toward building more resilient, scalable, and efficient network infrastructures compared to traditional cSDN [14].

In a dSDN architecture, each domain is equipped with its own local controller, responsible for building and maintaining a complete view of its local network state. These local controllers then synchronize their views with one another, ensuring that all controllers share a consistent and global perspective of the overall network [14]. Based on this global view, each local controller is capable of computing end-to-end paths and embedding the forwarding decisions directly into packet headers using strict source routing [14]. As packets traverse the network, the local controllers enforce these pre-computed routes embedded in the headers, guaranteeing that traffic flows are delivered efficiently and reliably to their destinations [14]. This design preserves the programmability and visibility of SDN while eliminating reliance on a single centralized controller.

Despite the advantages of dSDN in overcoming the limitations of centralized architectures, decentralized environments remain highly vulnerable to Distributed Denial of Service (DDoS) attacks. Similar to cSDN, attackers can overwhelm local controllers by flooding switches with malicious packets, leading to controller saturation, congestion of control–data plane links, and flow table exhaustion in switches [8,10, 15, 16]. However, dSDN introduces additional challenges that exacerbate the problem. Since each domain relies on local controllers that must synchronize to maintain a consistent global network view, DDoS attacks targeting one or more controllers can disrupt topology synchronization, resulting in outdated routing decisions and degraded inter-domain communication. Furthermore, the strict source-routing mechanism used in dSDN makes cross-domain forwarding dependent on cooperation among controllers; when a local controller is incapacitated, routing breakdowns and cascading failures can propagate across domains. Consequently, while decentralization reduces the risk of a single point of failure, it simultaneously expands the attack surface, making DDoS attacks a severe and complex threat to the resilience and stability of dSDN infrastructures.

Despite the extensive body of research on DDoS detection in SDN, existing solutions reveal several critical shortcomings. The first limitation is that almost all prior studies are designed for cSDN, while dSDN remains unexplored. The second issue lies in the dependence on supervised machine learning and deep learning models, which demand large, labeled datasets and continuous retraining, making them impractical for dynamic and evolving attack environments. Even with the rise of LLMs, prior work still relies heavily on fine-tuning or task-specific training, which requires significant computational resources and limits adaptability. Third, existing methods typically depend on complex flow-level statistics that are computationally expensive to generate and only available after flows are fully established, introducing delays unsuitable for rapid attack response. Finally, prior work offers little support for early and source-based detection, since attacks are usually identified only after traffic aggregation. Together, these limitations highlight a pressing need for new methods that are lightweight, adaptive, and capable of operating effectively in dSDN without the overhead of training or fine-tuning, and without reliance on complex flow feature extraction.

To address the limitations identified in prior studies and strengthen the security of dSDN, this study proposes a novel approach that combines architectural innovation with the power of LLMs. The key contributions of this work can be summarized as follows:

1. First DDoS detection and mitigation framework for dSDN: This work is among the first to propose and evaluate a dedicated DDoS detection and mitigation system tailored for dSDN, a recently emerging paradigm that addresses the limitations of traditional cSDN. The proposed framework demonstrates how security mechanisms can be effectively integrated into this new architecture.

2. Zero-training detection with the pre-trained DeepSeek-v3 model: We introduce a novel detection strategy that employs a pre-trained LLM (DeepSeek-v3) without the need for fine-tuning or additional training. By leveraging prompt-based in-context learning, the model dynamically adapts to traffic behavior and accurately identifies DDoS patterns, thereby eliminating the retraining overhead commonly required.

3. Lightweight port-level feature representation for early and source-based detection: Unlike traditional methods that depend on complex and resource-intensive flow-level statistics, our system relies on port-level features that are simpler, faster to extract, and more suitable for distributed controllers. This enables efficient detection of attacks directly at the source port and allows mitigation to be applied immediately, preventing propagation across the network.

In summary, this study pioneers the integration of pre-trained LLMs with port-level statistics in dSDN, offering a scalable, adaptive, and training-free solution for DDoS detection and mitigation.

To structure the remainder of this paper, Section 2 reviews related work on DDoS detection in SDN environments and identifies the research gaps. Section 3 presents the methodology, including the implementation of the dSDN architecture and the design of the proposed detection and mitigation system. Section 4 reports the experimental setup, test dataset construction, evaluation metrics, and results, followed by comparative analyses under different attack scenarios and against alternative LLMs and existing studies. Finally, Section 5 concludes the paper and outlines potential directions for future research.

## 2. Related Work

Over the past decade, SDN has attracted extensive research attention due to its flexibility and programmability, but at the same time, it has become a prime target for DDoS attacks [17]. Consequently, a wide range of defense mechanisms has been proposed, spanning from traditional traffic monitoring approaches to more advanced techniques based on machine learning, deep learning, and, more recently, fine-tuning of LLMs. These studies have focused primarily on cSDN architectures, investigating how statistical features, flow-level measurements, and traffic behavior can be exploited to distinguish normal traffic from malicious activities. A systematic review of this literature is essential to understand the state of the art, identify dominant methodologies and their limitations, and provide the basis for positioning our proposed work within the broader context of dSDN security research. Table 1 summarizes representative studies on DDoS detection in SDN, outlining their models, Datasets, methodologies, and main Findings. This overview provides the foundation for identifying critical gaps that remain unaddressed in the existing literature.

Table 1: Summary of the existing literature.

| Study | Model | Dataset | Methodology | Main Findings |
|---|---|---|---|---|
| [18] | BERT+ RF, DNN, CNN | InSDN | Data preprocessing (removing socket features), Random Forest feature selection (top 10 features), transforming flows into NLP sentences, multi-flow combination (4 flows), fine-tuning BERT-base-uncased, compared with CNN/DNN, evaluated on two scenarios (known and unseen attacks) | Achieved 99.96% accuracy, precision, recall, and F1 in both known and unseen attack detection; robust against zero-day/unseen attacks; outperformed CNN and DNN baselines |
| [19] | BRS + CNN | CICDDoS2019 | Data preprocessing, Balanced Random Sampling to balance the dataset, feature selection (Info Gain), CNN training, Mitigation using filtering, rate limiting, and iptables, tested in Mininet/POX | Achieved 99.99% (binary) and 98.64% (multi-class); effective mitigation; email notification; outperformed prior works |
| [20] | RDAER | CICDDoS2019 | Data preprocessing, feature selection (RFE), traffic clustering (DBSCAN), anomaly prediction (ARIMA, Lyapunov exponent, exponential smoothing, dynamic threshold), and event correlation for detection | Achieved 99.92% accuracy; early detection at switch level; reduced false positives and resource usage compared to prior works |
| [21] | DNN | InSDN, CICIDS2018, Kaggle DDoS | Data preprocessing, custom DNN architecture, hyperparameter tuning, real-time detection & mitigation integrated into the Ryu SDN controller | Achieved 99.98%, 100%, and 99.99% accuracy on the three datasets, demonstrating effective real-time mitigation in an emulated SDN environment |
| [22] | AE-BGRU | Custom datasets | Feature extraction (IP headers, ToS, inter-arrival time, unknown destination addresses, switch capacity, flows, IP options), DL detection with AE-BGRU, trust-value mechanism for mitigation, simulated in Mininet with POX controller, TensorFlow implementation | Achieved 99.91% accuracy (data plane) and 99.89% (control plane); low false positive rate (0.09%); mitigation via dynamic trust value and blocking; outperformed existing ML/DL methods |
| [23] | Hybrid CNN-LSTM with XGBoost | CICDDoS2019 | Used XGBoost for feature selection and a hybrid CNN-LSTM model for classification in SDN-based IIoT networks | Achieved 99.50% accuracy; very low latency; lightweight architecture; robust binary and multi-class classification; outperformed existing ML/DL models |
| [24] | SVM, RF, KNN, XGBoost, NB | Custom datasets | Created SDN topology in Mininet; generated normal + attack traffic using *hping3*; captured flow stats in Ryu; extracted 7 features (flow count diff, byte count diff, packet count diff, source IP count, ratios, etc.); trained and tested classifiers; best (SVM) deployed in Ryu controller. | SVM achieved 99.4% accuracy, precision, recall, F1; AUC 0.995; FAR 0.72%; better than RF, KNN, NB, XGBoost; outperformed prior works |
| [25] | DNN | CICDDoS2019 | Data preprocessing (removing irrelevant/misleading features, normalization), three-layer feedforward deep neural network trained with AdaMax optimizer | Achieved 99.99% accuracy for binary detection and 94.57% for multi-class classification; reliable detection; suitable for IDS integration |
| [26] | SVC, KNN, RF, ANN, LR | Custom dataset | Feature engineering; dataset creation in Mininet; flow & port statistics logging; machine learning classifiers applied; hybrid Support Vector Classifier with Random Forest for classification | Hybrid SVC-RF achieved the best performance, reaching 98.8% accuracy, 97.91% detection rate, 98.18% specificity, and the lowest false alarm rate (0.02%). The dataset was made public, and the model demonstrated strong performance across TCP, UDP, and ICMP protocols |
| [27] | SVM, NB, ANN, KNN + Feature Selection (Relief, SFS, Lasso) | Custom dataset | Data collected in Mininet/POX SDN with sFlow + InfluxDB; applied feature selection (filter, wrapper, embedded); trained/tested with SVM, NB, ANN, KNN classifiers | Wrapper + KNN achieved highest accuracy (98.3%); feature selection reduced processing load and improved efficiency |
| [28] | ASVM | Custom dataset | Traffic generation (normal, UDP flood, SYN flood), data collection via OpenFlow switches, feature extraction (volumetric & asymmetric features: ANPI, ANBI, VPI, VBI, ADTI), classification with ASVM (linear kernel, OVS decision function), evaluation using cross-validation | Achieved 97% detection accuracy; reduced training and testing time compared to standard SVM; effective detection of SYN and UDP flooding attacks |

| [29] | SVM | Custom dataset | Extracted 6-tuple flow features (IP speed, port speed, flow packet/byte deviation, flow entries, pair-flow ratio); built and trained SVM classifier; evaluated on TCP, UDP, ICMP flood attacks using Hping3 | Achieved 95.24% average detection accuracy, 1.26% false alarm rate; effective across multiple DDoS types (TCP, UDP, ICMP) |

The majority of prior work has concentrated on cSDN architectures, leaving the emerging decentralized paradigm largely unexplored. While decentralization addresses critical limitations such as scalability, complexity, and single points of failure, it also introduces unique challenges, including inter-controller synchronization and cross-domain routing. The absence of dedicated defense mechanisms tailored for dSDN means that this architecture remains highly vulnerable to DDoS attacks, highlighting an urgent need for solutions that specifically address its distributed nature.

Previous approaches have largely relied on supervised machine learning, deep learning, and, more recently, fine-tuning of LLMs. These methods require large volumes of labeled data, frequent retraining, and task-specific adjustments, all of which introduce significant computational and operational costs. While such techniques have demonstrated promising accuracy in controlled environments, their dependency on pre-collected datasets makes them less adaptive to dynamic and evolving attack strategies. In practice, gathering and labeling sufficient data for every new type of attack is costly and time-consuming, leaving networks exposed to emerging threats. This overreliance on supervised training and fine-tuning thus limits the practicality and resilience of current solutions when applied to real-world and dSDN environments.

A common limitation observed in existing studies is their dependence on complex flow-level features, which require significant computational effort and can only be extracted once network flows are fully established. This reliance not only introduces delays in the detection process but also creates overhead that makes real-time or near-real-time defense challenging, especially in dSDN environments where multiple controllers operate simultaneously. Flow-level statistics, such as aggregated traffic patterns or per-flow counters, are resource-intensive to generate and maintain, leading to scalability issues as the network grows. Moreover, the latency incurred by waiting for complete flow information undermines the ability to detect attacks in their early stages. As a result, the dependence on flow-level features limits both the efficiency and responsiveness of existing DDoS defense mechanisms, leaving dSDN infrastructures particularly vulnerable to fast-spreading attacks.

Another persistent shortcoming is the lack of early and source-based detection mechanisms. Most prior methods identify DDoS attacks only after traffic flows have been aggregated and analyzed, which significantly delays the response and allows malicious traffic to propagate through the network. In dSDN environments, this delay is even more problematic, as attack traffic may quickly spread across multiple domains before being recognized, making mitigation far more complex. Without detection at the source ports, controllers are left to react only after the attack has already consumed valuable network resources, reducing both effectiveness and timeliness. The absence of early, source-aware detection therefore represents a critical shortcoming, as it prevents proactive defense and leaves dSDN infrastructures exposed to cascading disruptions caused by rapidly evolving DDoS attacks.

Taken together, these research gaps underscore the pressing need for novel approaches that are lightweight, adaptive, and capable of operating effectively in dSDN environments.

3. Methodology

The methodology of this study is designed to translate the identified research gaps into a concrete framework for securing dSDN against DDoS attacks. While existing works largely depend on flow-level statistics, supervised learning, or fine-tuned LLMs, our approach emphasizes lightweight features, training-free adaptability, and early source-based detection. To this end, the methodology unfolds in three stages. First, we implement a dSDN testbed using open-source platforms to provide a realistic and controllable environment for experimentation. Second, we design a detection pipeline that leverages port-level feature extraction and prompt engineering to enable in-context learning with the pre-trained DeepSeek-v3 model, avoiding the overhead of training or fine-tuning. Finally, we integrate detection with automated mitigation at the local controller level, ensuring that malicious traffic is immediately suppressed at its source before it propagates across domains. Collectively, these methodological components establish a distributed, proactive, and scalable defense system tailored to the unique operational requirements of dSDN.

## 3.1 Implementation of Decentralized SDN (dSDN)

To realize a practical dSDN environment, we adopted the architectural principles proposed in prior work on multi-controller synchronization and strict source routing [14]. However, since no dedicated simulation frameworks currently exist for dSDN, we implemented the architecture using open-source tools, namely Mininet for network emulation and the Ryu controller platform for programmability. This implementation enables the creation of multiple local controllers distributed across different domains, each managing its own switches and network resources. The experimental setup provides a realistic and controllable testbed that supports both intra-domain and inter-domain communication, serving as the foundation for integrating and evaluating our proposed DDoS detection and mitigation system.

Each domain within the dSDN environment is equipped with its own local controller, responsible for maintaining a comprehensive view of the domain's topology, traffic statistics, and forwarding rules. To avoid isolated decision-making and ensure consistent global behavior, these local controllers periodically synchronize their state information with one another. Through this synchronization process, every controller obtains not only a local perspective but also an up-to-date global view of the overall network. This distributed yet coordinated control plane eliminates reliance on a single centralized controller while preserving the logical consistency required for coherent network management across domains.

Based on the synchronized global view, each local controller is capable of computing optimal end-to-end paths for network traffic. Instead of relying on intermediate routing decisions at every switch, the controllers embed the complete forwarding path directly into packet headers using strict source routing. In our implementation, this functionality is realized through Multiprotocol Label Switching (MPLS), where labels are inserted into packet headers to represent the precomputed routes. As packets traverse the network, local controllers and switches forward them strictly according to these MPLS labels, ensuring that flows follow their assigned paths until reaching their destinations. By shifting routing intelligence into the header information through MPLS, the system reduces latency, minimizes control-plane overhead, and improves the reliability of inter-domain communication. The overall architecture of the implemented dSDN environment, including local controllers, inter-controller synchronization, and MPLS-based source routing, is illustrated in Figure 1.

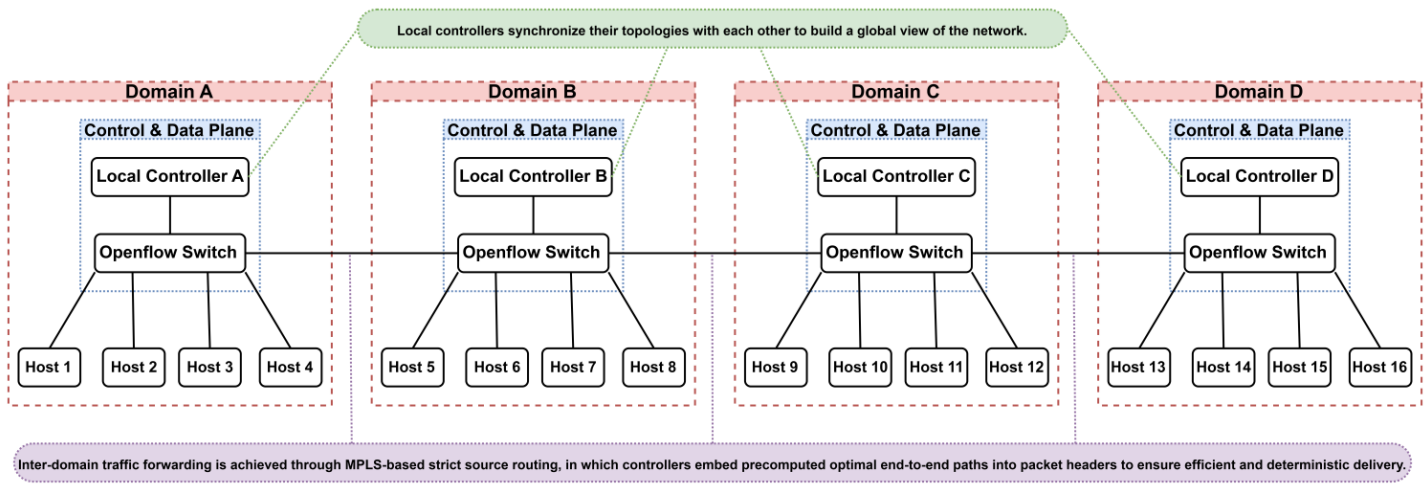

Figure 1: dSDN architecture with multiple local controllers. Each controller manages its domain, synchronizes with peers to maintain a global view, and forwards packets across domains using MPLS-based strict source routing.

The dSDN architecture implemented in this study not only facilitates scalable and reliable routing but also provides a foundation for integrating security mechanisms. Since each local controller maintains both a local and synchronized global network view, it can observe traffic patterns at the port level and detect anomalies close to their source. By embedding strict source routing information using MPLS labels, controllers ensure that forwarding decisions are deterministic, which simplifies monitoring and detection of deviations from expected traffic behavior. This tight integration between routing and security enables the controllers to serve a dual role: forwarding legitimate flows while simultaneously acting as distributed detection and mitigation points. The next section builds upon this foundation by introducing our proposed detection and mitigation framework, which leverages port-level features and LLMs to identify and counter DDoS attacks in dSDN environments.

## 3.2 Proposed Detection and Mitigation System

To complement the dSDN architecture and address its vulnerability to DDoS attacks, we design a lightweight yet effective detection and mitigation framework. The proposed system operates directly within local controllers, leveraging port-level statistics for efficient feature extraction and integrating a pre-trained LLM (DeepSeek-v3) through prompt-based in-context learning. This design enables the system to analyze traffic behavior dynamically without any additional training overhead. Moreover, by embedding detection and mitigation capabilities at the source ports, the framework provides proactive protection, ensuring that malicious traffic is contained before it propagates across domains. The following subsections describe each component of the proposed system in detail.

### 3.2.1 Port-level Feature Extraction

In this study, we relied on port-level statistics as the primary features for traffic analysis. Specifically, the system collects four lightweight counters at each switch port: the number of received packets, the number of received bytes, the number of transmitted packets, and the number of transmitted bytes. These statistics are readily available from the OpenFlow interface and can be retrieved with minimal computational overhead. In our implementation, these counters are periodically aggregated every 10 seconds, providing a time-windowed representation of traffic dynamics that balances responsiveness with stability.

We deliberately adopt port-level features instead of traditional flow-level features for several reasons. First, flow-level statistics can only be generated after a flow is fully established, which introduces delays in detection and makes them unsuitable for early response. In contrast, port-level features are available instantly at switch interfaces, enabling continuous monitoring and early detection of abnormal traffic patterns. Second, extracting and aggregating flow-level data requires significant processing and memory resources, which increases the computational burden on controllers and may limit scalability. Port-level features, however, are lightweight, faster to extract, and impose minimal overhead, making them more practical for distributed detection systems in dSDN environments. Finally, port-level monitoring provides direct visibility into source behavior, allowing attacks to be detected and mitigated directly at the source ports, whereas flow-level analysis typically identifies attacks only after traffic aggregation has already occurred.

By focusing on these simple yet effective port-level counters, aggregated in 10-second intervals, our system ensures fast, scalable, and distributed monitoring, which is crucial for achieving proactive defense in dSDN environments.

### 3.2.2 Prompt Engineering with In-Context Learning and Zero-Training with DeepSeek-v3

A central element of our framework is the use of prompt engineering to enable in-context learning, a paradigm where LLMs adapt to new tasks by conditioning on examples provided directly within the prompt, without modifying their parameters [30]. During each monitoring window, the aggregated port-level features, namely the counts of received packets, received bytes, transmitted packets, and transmitted bytes, are transformed into natural language prompts that describe the observed traffic behavior. To guide the model's reasoning, these prompts are augmented with ten labeled examples of benign traffic only, chosen because benign examples are more stable and representative of normal operation, whereas attack traffic is highly variable. By embedding such examples directly into the input, the system enables the LLM to dynamically infer whether new traffic patterns correspond to normal activity or potential DDoS attacks. This approach shifts the adaptation process from retraining the model to carefully constructing prompts, making it possible for the LLM to respond to evolving traffic behaviors without the need for costly parameter updates. An illustration of this process is shown in Figure 2, which provides an example of the prompt automatically generated by the system during execution for each port.

> [Task]: Detect whether the interface status over the last ten seconds represents a flood-based DDoS attack (1) or Normal traffic (0). Analyze the labeled examples provided, then classify the new interface status accordingly.
> [Labeled interface status]:
> - (Received 1314 packets with 1923532 bytes and sent 1314 packets with 86736 bytes) => 0
> - (Received 1031 packets with 1508646 bytes and sent 1030 packets with 67980 bytes) => 0
> - (Received 4329 packets with 6337922 bytes and sent 4326 packets with 285524 bytes) => 0
> - (Received 2224 packets with 3257592 bytes and sent 2223 packets with 146726 bytes) => 0
> - (Received 4690 packets with 6864356 bytes and sent 4686 packets with 309292 bytes) => 0
> - (Received 834 packets with 1201256 bytes and sent 2 packets with 273 bytes) => 0
> - (Received 1510 packets with 2177420 bytes and sent 1 packets with 1512 bytes) => 0
> - (Received 2343 packets with 3378606 bytes and sent 1 packets with 1512 bytes) => 0
> - (Received 2972 packets with 4285624 bytes and sent 0 packets with 0 bytes) => 0
> - (Received 3276 packets with 4723992 bytes and sent 1 packets with 1512 bytes) => 0
> [New interface status]:
> - (Received 1361 packets with 1953154 bytes and sent 9 packets with 2268 bytes) => ???
> [Instruction]: Only answer with one number: 0 if Normal, or 1 if DDoS. Do not explain.

Figure 2: Example of our generated prompt for DeepSeek-v3.

To realize this design, we employ DeepSeek-v3 [31], a pre-trained LLM, used directly in a zero-training configuration without any fine-tuning or retraining on task-specific datasets. Unlike prior studies that rely on supervised machine learning, deep learning, or fine-tuned LLMs, our approach leverages DeepSeek-v3's extensive pre-existing knowledge and adapts to new traffic conditions solely through the contextual information provided in the prompt. The zero-training paradigm eliminates the overhead of dataset labeling, parameter optimization, and repeated training cycles, all of which are impractical in dynamic and dSDN environments. By avoiding these burdens, our framework achieves faster deployment, reduced computational cost, and greater scalability, making it well-suited for real-time DDoS detection and mitigation in dSDN infrastructures.

### 3.2.3 Integrated Early Detection and Mitigation

To achieve effective protection in dSDN environments, it is not sufficient to detect anomalies in isolation; detection must be tightly integrated with mechanisms for immediate mitigation. In our framework, this integration is realized through the integration of port-level monitoring, prompt-based inference with DeepSeek-v3, and automated controller responses into a unified pipeline. Every 10 seconds, port-level features of each host's port are collected and converted into natural language prompts enriched with benign examples for in-context learning, which allow the LLM to classify the traffic status without retraining. Once an anomaly is flagged as a potential DDoS attack, the local controller immediately initiates mitigation by installing drop rules on the attacker's port. This design ensures that malicious traffic is contained at its point of origin, preventing it from consuming network resources or propagating across domains. By embedding both detection and mitigation within the control logic of each local controller, the framework provides a distributed, proactive, and low-latency defense strategy tailored to the characteristics of dSDN.

Early and source-based detection is a central capability of our framework, distinguishing it from prior approaches that typically rely on flow-level aggregation. By operating directly on lightweight port-level statistics features, the system can identify abnormal traffic patterns as soon as they emerge at the attacker's port. This enables local controllers to detect DDoS attacks during their initial stages, before malicious flows are fully established or propagated across domains. The ability to localize detection at the attacker's port ensures that anomalies are captured with minimal delay, reducing the risk of network-wide disruptions. In the context of dSDN, where multiple local controllers coordinate to maintain a global view, such localized detection provides a significant advantage by confining threats within their originating domain and preventing cascading failures in inter-domain communication.

Once a DDoS attack is detected, the mitigation process is triggered immediately at the local controller. This is accomplished by dynamically installing drop rules on the attacker's port, ensuring that malicious packets are discarded at the point of origin. This proactive intervention prevents attack traffic from consuming bandwidth, overloading switches, or propagating into other domains. To avoid unnecessary long-term disruption of legitimate services, the system incorporates an automatic recovery mechanism: after 30 seconds of attack inactivity, the affected port is re-enabled and resumes normal operation. If malicious traffic reappears, the detection and mitigation cycle is automatically reactivated, thereby ensuring continuous protection. Because the mitigation is executed locally and autonomously by each local controller, the framework achieves low-latency response and distributed defense across all domains. This design ensures that even if multiple attacks occur simultaneously in different parts of the network, each domain can independently detect, suppress, and recover from malicious flows without relying on centralized coordination.

The integration of detection and mitigation within each local controller offers several key advantages over conventional approaches. First, the system provides proactive defense by detecting and neutralizing DDoS attacks at their source ports before they propagate, thereby minimizing their impact on the broader network. Second, the use of lightweight port-level features and prompt-based inference ensures low-latency operation, enabling the framework to react within seconds rather than waiting for flow-level aggregation. Third, the decentralized design inherently supports distributed mitigation, as each local controller independently monitors, classifies, and suppresses malicious traffic within its own domain. Together, these characteristics result in a defense mechanism that is adaptive, scalable, and well-aligned with the operational requirements of dSDN environments.

To provide a clear overview of how the proposed framework operates in practice, Figure 3 illustrates the workflow of the proposed framework, from port-level feature collection and prompt generation to traffic classification with DeepSeek-v3 and automated mitigation at the attacker's port. This flowchart emphasizes the seamless integration of detection and mitigation within each local controller.

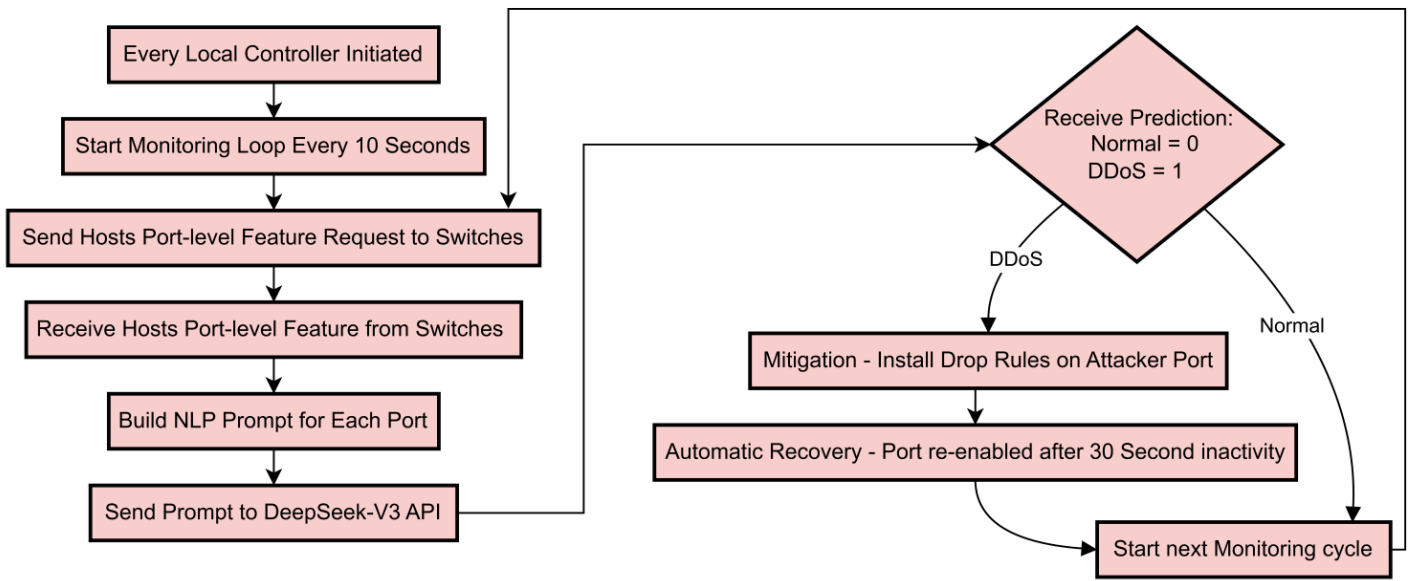

Figure 3: Flowchart of the proposed detection and mitigation framework in dSDN, showing the pipeline from traffic monitoring to classification and automated response.

## 4. Results and Discussion

This section presents the experimental evaluation of the proposed DDoS detection and mitigation framework in dSDN. The results are organized to demonstrate the effectiveness of the system across multiple dimensions, starting with the description of the experimental setup and the test dataset, followed by the performance metrics used for evaluation. We then report the detection accuracy of the DeepSeek-v3-based model, analyze its behavior under different flood attack scenarios, and compare its performance against other LLMs as well as existing studies in the literature. Through this systematic analysis, we aim to validate the practicality, scalability, and robustness of the proposed approach in addressing DDoS threats in dSDN environments.

## 4.1 Experimental Setup Environment

The experimental evaluation was conducted on a dSDN testbed built using widely adopted open-source tools. Mininet was used to emulate the network topology, while the Ryu controller platform was employed to manage the control plane and enforce routing and mitigation rules. For traffic generation, the iperf tool was used to create normal traffic, whereas hping3 was employed to simulate DDoS flood attacks, including ICMP, UDP, and TCP floods. All experiments were performed on a server running Ubuntu 20.04.6 LTS with Linux kernel 5.15.0-139-generic, equipped with dual Intel Xeon Gold 6230R processors (26 cores per socket, 52 physical cores / 104 threads total) clocked at 2.10 GHz, and 1.2 TiB of RAM. This setup provided both a realistic emulation environment for dSDN and the computational resources necessary for real-time inference with the DeepSeek-v3 model. Table 2 summarizes the tools, traffic generators, and hardware specifications used in our experimental setup.

Table 2: Experimental setup: tools, traffic generators, and hardware specifications.

| Component | Description |
| --- | --- |
| Network Emulator | Mininet (to emulate dSDN topology with multiple local controllers) |
| Controller Platform | Ryu controller (Python-based, extended for multi-local-controller operation) |
| Traffic Generator | iPerf (to generate normal traffic across domains). |
| Attack Tool | hping3 (to launch DDoS floods: ICMP, UDP, and TCP). |
| Server Hardware | Ubuntu 20.04.6 LTS (Linux kernel 5.15.0-139-generic); Dual Intel Xeon Gold 6230R (26 cores per socket, 52 physical cores / 104 threads total) @ 2.10 GHz; 1.2 TiB RAM. |

## 4.2 Traffic Generated and Testing Dataset

To generate realistic experimental traffic, we designed a dSDN topology consisting of four domains, where each domain contained one local controller connected to a single switch, and each switch was further connected to four hosts, as illustrated in Figure 4. Within this topology, three hosts were configured as dedicated servers: h5 operated as an HTTP server, h6 as a TCP iperf server, and h7 as a UDP iperf server. The remaining hosts functioned as clients, continuously generating normal background traffic toward the servers. To emulate realistic communication dynamics, client hosts were programmed to initiate diverse connections, including ICMP pings, TCP flows, UDP flows, and HTTP requests, with both the target servers and the client initiators selected randomly. This setup ensured that the resulting traffic captured a broad spectrum of benign behaviors, thereby reflecting the heterogeneity and unpredictability of real-world network environments.

To evaluate the resilience of the proposed framework under adversarial conditions, we introduced three distinct DDoS attack scenarios with varying traffic rates: ICMP flood, TCP flood, and UDP flood. For each attack type, multiple client hosts from different domains were designated as attackers, thereby reflecting the distributed nature of botnet-driven DDoS campaigns, as illustrated in Figure 4. The attacking hosts generated high-rate malicious traffic directed at the designated servers, where h5 was targeted by ICMP floods, h6 by TCP floods, and h7 by UDP floods, with the goal of overwhelming the services with excessive requests. Each attack scenario was executed independently for a continuous duration of two hours to isolate its effects and ensure controlled evaluation. During execution, port-level statistics were continuously collected and converted into natural language prompts, while ground-truth labels were systematically assigned based on the timing and source of the attacks. This methodology enabled the test dataset to capture both benign and malicious traffic patterns in a realistic yet controlled setting, providing a robust foundation for assessing the detection and mitigation capabilities of the proposed framework.

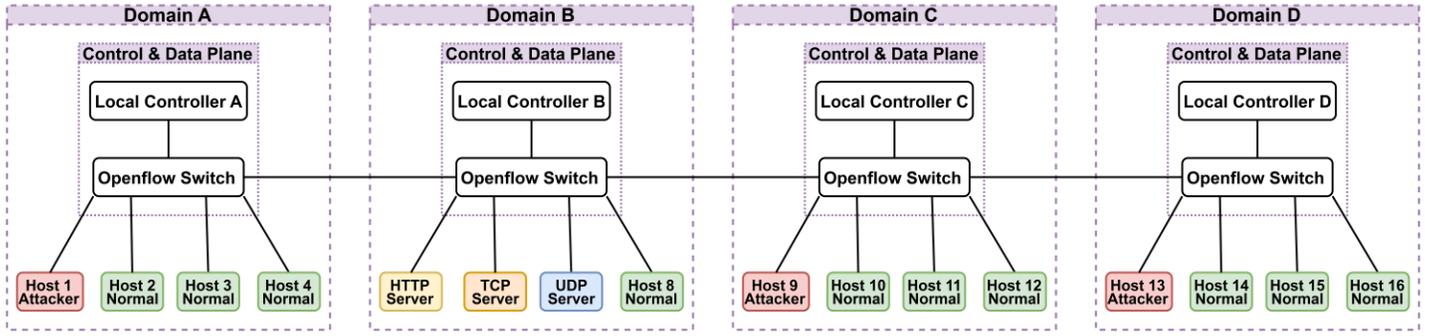

Figure 4: dSDN topology, HTTP server, TCP server, UDP server, normal hosts, and attackers.

During the execution of both normal traffic and attack scenarios, the system continuously aggregated port-level features into natural language prompts at 10-second intervals. Over the course of all experiments, this process yielded a total of 25,933 prompts, each representing the traffic state of a specific port within a defined time window. For every prompt, the deepseek-v3 model's predictions were stored alongside ground-truth labels, which were assigned according to the known timing and origin of the attacks. All prompts, predictions, and labels were systematically archived into a CSV test dataset, ensuring reproducibility and enabling detailed quantitative analysis. The distribution of prompts across normal and attack traffic is summarized in Table 3, which highlights benign and adversarial scenarios captured in the dataset.

Table 3: Distribution of prompts in the experimental test dataset.

| Traffic Type | Number of Prompts |
|---|---|
| Normal | 19,504 |
| TCP Flood | 2,151 |
| UDP Flood | 2,132 |
| ICMP Flood | 2,146 |
| Total | 25,933 |

### 4.3 Performance Metrics

To rigorously evaluate the performance of the proposed detection and mitigation framework, we employ a set of widely adopted classification metrics. These metrics quantify different aspects of predictive performance, including overall accuracy, sensitivity to attacks, and balance between false alarms and missed detections. Let *TP, TN, FP,* and *FN* denote true positives, true negatives, false positives, and false negatives, respectively.

- Accuracy (ACC): measures the proportion of correctly classified instances among all instances:

$$Accuracy = \frac{TP + TN}{TP + TN + FP + FN} \quad (1)$$

- Precision (PRE): quantifies the proportion of predicted attacks that are truly attacks, reflecting resistance to false alarms:

$$Precision = \frac{TP}{TP + FP} \quad (2)$$

- Recall (REC): also known as detection rate or sensitivity, measures the proportion of actual attacks that are correctly identified:

$$Recall = \frac{TP}{TP + FN} \quad (3)$$

- F1-Score (F1): provides the harmonic mean of precision and recall, offering a balanced view when both false positives and false negatives are critical:

$$F1 - Score = 2 * \frac{Precision * Recall}{Precision + Recall} \quad (4)$$

- Confusion Matrix: a tabular representation that summarizes the distribution of true and predicted labels across the classes, enabling detailed inspection of misclassification patterns.
- Receiver Operating Characteristic (ROC) and Area Under the Curve (AUC): the ROC curve plots the true positive rate (TPR) against the false positive rate (FPR) at varying thresholds, while the AUC summarizes this curve into a single scalar value that reflects the overall discriminative ability of the model.

Together, these metrics provide a comprehensive evaluation of the detection system, ensuring both effectiveness in identifying attacks and reliability in minimizing false alarms.

### 4.4 Results of the Proposed DeepSeek-v3 Model

To evaluate the effectiveness of the proposed framework, we relied on the dataset generated during six hours of continuous system execution, comprising two hours for each attack scenario (ICMP, UDP, and TCP floods). Throughout these experiments, port-level features were aggregated every 10 seconds, transformed into prompts, and classified by the DeepSeek-v3 model in its zero-training configuration. The corresponding predictions were stored alongside ground-truth labels, which were assigned based on the exact timing and source of each attack, thereby ensuring accurate evaluation. The analysis focused on the metrics introduced in Section 4.3, namely accuracy, precision, recall, F1-score, confusion matrix, and the Area Under the ROC Curve (AUC). These metrics collectively provide a comprehensive view of the model's ability to correctly distinguish between normal and attack traffic while minimizing false alarms. The results demonstrate that DeepSeek-v3 can successfully adapt to the detection task through prompt engineering and in-context learning, achieving strong detection performance without the need for additional training.

The experimental results of the proposed DeepSeek-v3 model reveal outstanding performance across all evaluation metrics. The model achieved an accuracy of 99.99%, precision of 99.97%, recall of 100%, and an F1-score of 99.98%. The confusion matrix presented in Figure 5 confirms these results, demonstrating that out of 25,933 prompts, only two benign instances were misclassified as attacks, while no attack traffic was misclassified as benign. This indicates an exceptionally low false-alarm rate, while maintaining perfect sensitivity in detecting attacks. Furthermore, the ROC curve illustrated in Figure 6 achieved an AUC of 1.00, highlighting the model's excellent ability to discriminate between normal and attack traffic. These results validate the effectiveness of combining port-level monitoring with prompt-based inference, enabling DeepSeek-v3 to provide near-perfect detection without the need for fine-tuning or retraining.

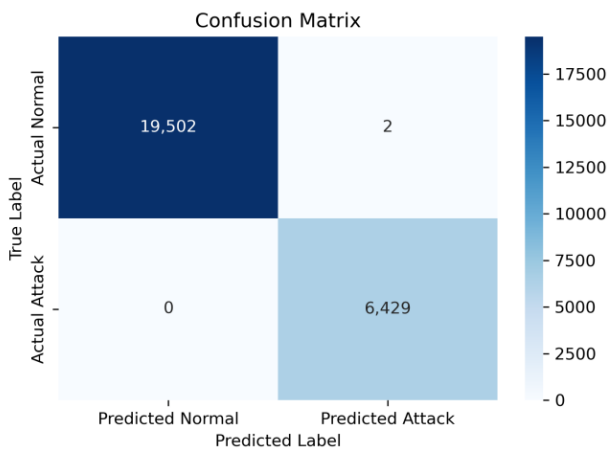
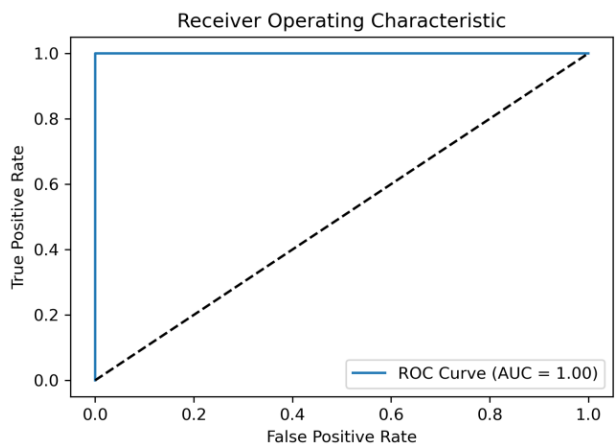

Figure 5: Confusion matrix of the proposed DeepSeek-v3 model     Figure 6: ROC-AUC curve of the proposed DeepSeek-v3 model

## 4.5 Real-time Evaluation under Different Flood Attack Scenarios

To further assess the robustness of the proposed framework, we conducted real-time experiments under three distinct flood-based DDoS attacks: TCP flood, UDP flood, and ICMP flood. Unlike the long-term dataset collection described earlier, these real-time experiments were observed over a 10-minute window (600 seconds), during which system performance was monitored at one-second intervals. Each attack scenario was executed independently and compared against both normal operation (no attack) and operation without mitigation. The evaluation focuses on four key aspects of system behavior: (i) the rate of Packet-in messages processed by local controllers, (ii) CPU utilization of local controllers, (iii) the stability of inter-controller synchronization, and (iv) the volume of incoming traffic received at the victim servers. These measurements collectively provide a comprehensive view of how the detection and mitigation pipeline performs in practical conditions, demonstrating its ability to contain malicious traffic, preserve controller stability, and protect network services across decentralized domains.

### 4.5.1 TCP Flood Attack Scenario

In the TCP flood scenario, multiple compromised hosts (h1, h9, and h13) from different domains simultaneously launched a high-rate flood of TCP connection requests targeting TCP server h6. The attack intensity was configured at 20,000 packets per second, reflecting a realistic large-scale flood capable of exhausting network resources. As shown in Figure 7, this caused a sharp surge in the number of Packet-in messages generated by the switches and forwarded to their respective controllers. Without mitigation, the Packet-in rate quickly escalated to thousands of messages per second, overwhelming controller processing capacity. Once the proposed framework was activated, however, the anomaly was detected within seconds at the source ports, and mitigation was enforced by installing drop rules. As a result, the Packet-in rate rapidly declined to baseline levels, demonstrating the framework's ability to suppress attack-induced overhead.

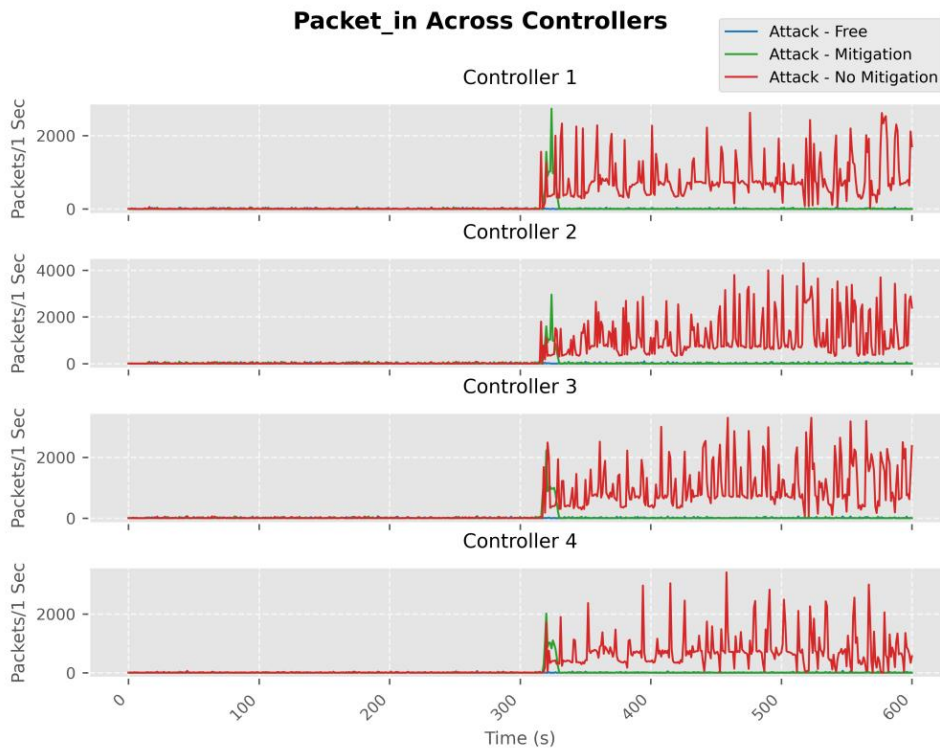

Figure 7: Packet_in across controllers in the TCP flood attack scenario

Similarly, Figure 8 illustrates the CPU utilization of local controllers. During the unmitigated attack, CPU usage spiked significantly due to excessive processing demands. After mitigation, CPU utilization stabilized close to its normal operating level, confirming that the framework prevents resource exhaustion by limiting malicious traffic at the source.

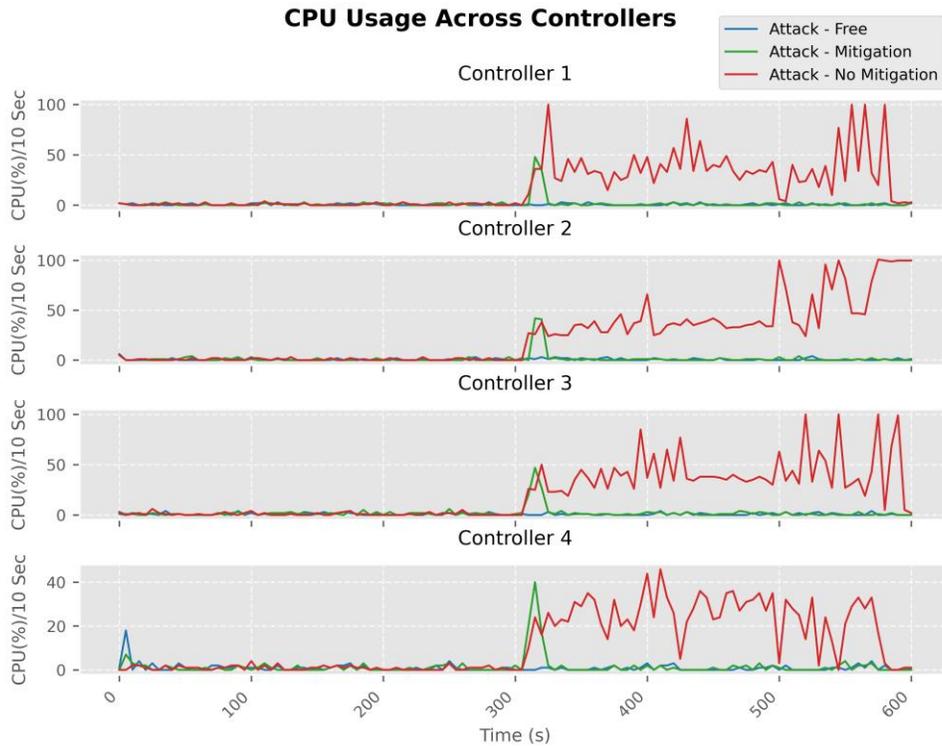

Figure 8: CPU usage across controllers in the TCP flood attack scenario

The effects of the TCP flood on inter-controller synchronization are depicted in Figure 9. Under attack conditions, synchronization delays increased due to congestion in control–data plane communications. With mitigation enabled, synchronization remained stable, ensuring consistent global network views across domains.

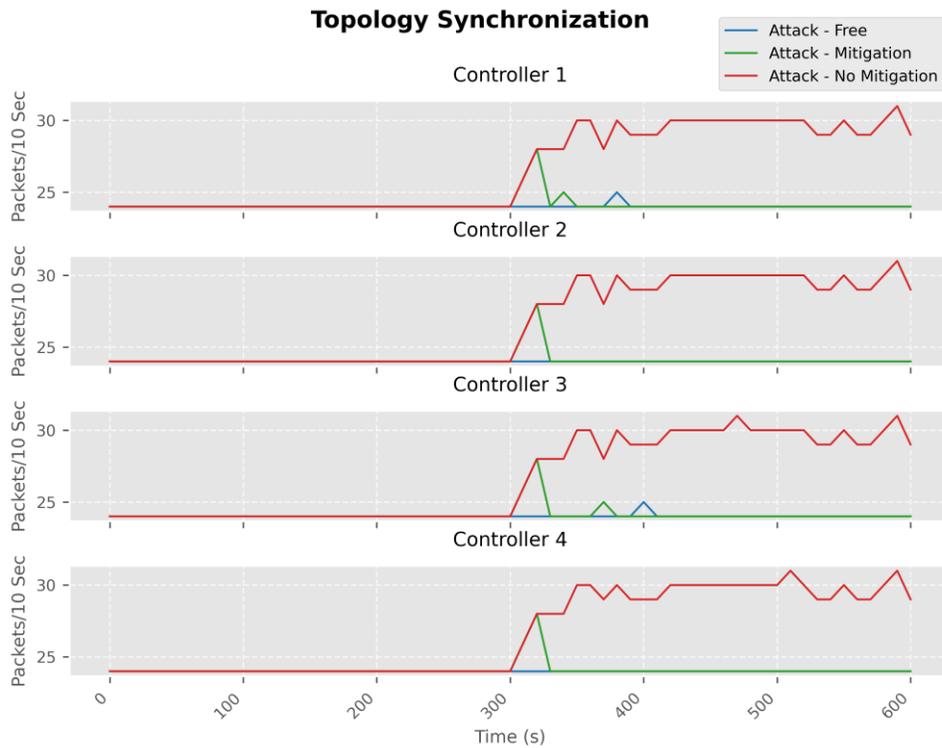

Figure 9: Peer topology synchronization in the TCP flood attack scenario

Finally, Figure 10 presents the incoming traffic received at the victim server (h6). While the unmitigated attack saturated the server with an unsustainable volume of TCP requests, mitigation restored traffic to normal levels, ensuring service continuity. Together, these results highlight the framework's ability to provide rapid, source-based defense against TCP flood attacks, preserving both controller stability and server availability in dSDN environments.

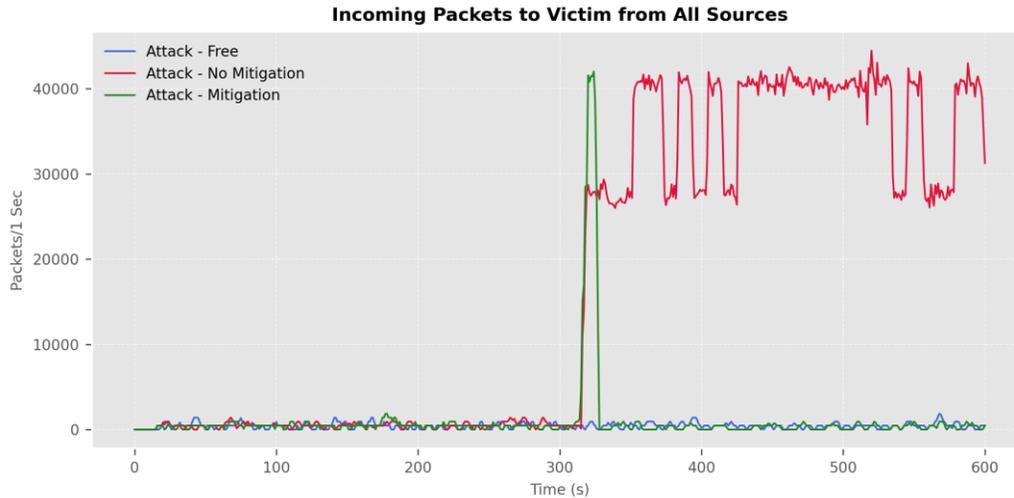

Figure 10: Incoming packets to the victim from all sources in the TCP flood attack scenario

### 4.5.2 UDP Flood Attack Scenario

In the UDP flood scenario, multiple attacking hosts (h1, h9, and h13) distributed across different domains generated a large-scale flood of UDP packets targeting the h7 UDP server. The attack rate was configured at 66,666 packets per second, representing a highly aggressive and resource-exhaustive traffic pattern. As illustrated in Figure 11, the attack caused an immediate and steep increase in Packet-in messages at the local controllers. Without defense, the Packet-in rate rose uncontrollably, threatening to saturate controller processing capacity and destabilize the control plane. With the proposed framework enabled, however, the anomaly was identified within seconds at the source ports, and drop rules were applied to suppress the malicious traffic. This intervention reduced the Packet-in rate back to near-baseline levels, preserving controller responsiveness.

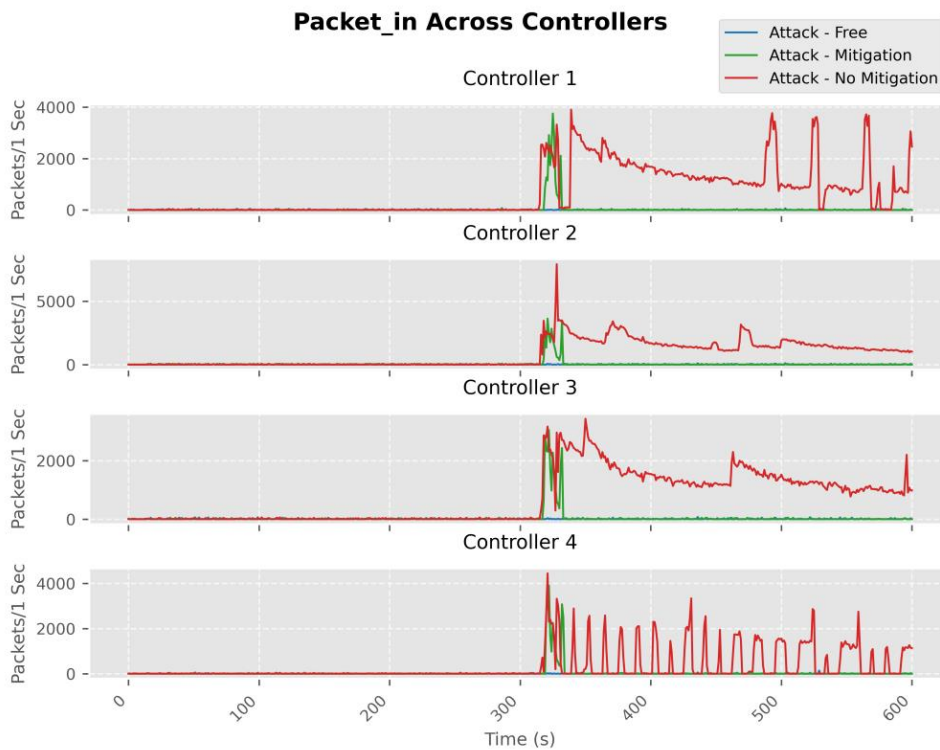

Figure 11: Packet_in across controllers in the UDP flood attack scenario

Figure 12 further demonstrates the impact on CPU utilization at local controllers. During the unmitigated UDP flood, CPU usage spiked sharply as controllers attempted to process the massive influx of control messages. Following mitigation, CPU utilization stabilized, remaining close to normal operating conditions, thereby confirming the system's effectiveness in alleviating resource exhaustion.

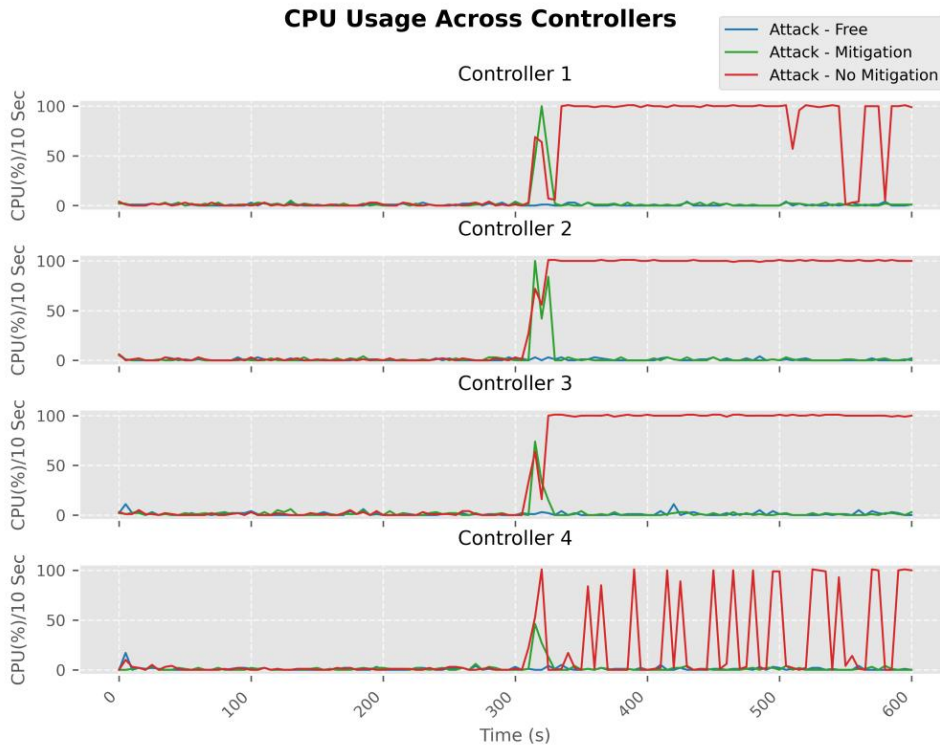

Figure 12: CPU usage across controllers in the UDP flood attack scenario

Unlike the TCP flood scenario, the UDP flood attack exhibited minimal impact on inter-controller synchronization, as illustrated in Figure 13. During the attack, synchronization delays remained largely stable, indicating that the controllers were able to maintain a consistent global view of the network. This behavior can be attributed to the characteristics of UDP traffic, where the attack primarily stressed packet forwarding and processing capacity rather than imposing significant overhead on synchronization exchanges. These results highlight that different types of flood-based attacks can exert distinct pressures on dSDN infrastructures, underlining the importance of evaluating multiple attack vectors when assessing system resilience.

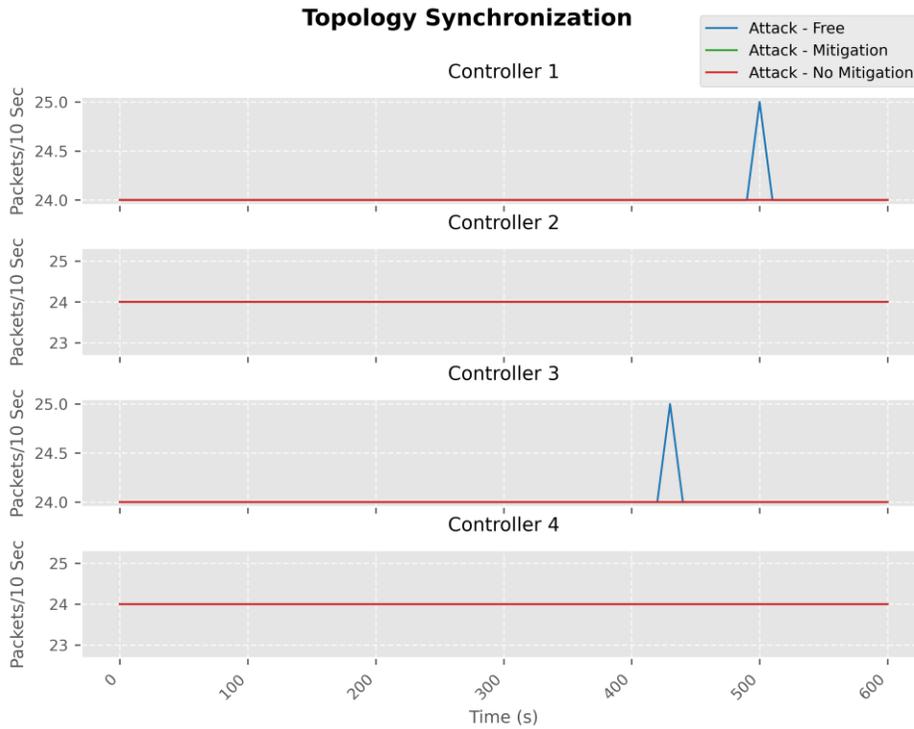

Figure 13: Peer topology synchronization in the UDP flood attack scenario

Lastly, Figure 14 shows the traffic received at the victim server (h7). Under the unmitigated attack, the server was overwhelmed by excessive UDP packets, rendering the service unusable. With the defense mechanism activated, incoming traffic was reduced to normal levels, allowing the server to remain fully operational. These results emphasize the framework's ability to rapidly neutralize high-speed UDP flood attacks at their source, ensuring stability and availability in dSDN environments.

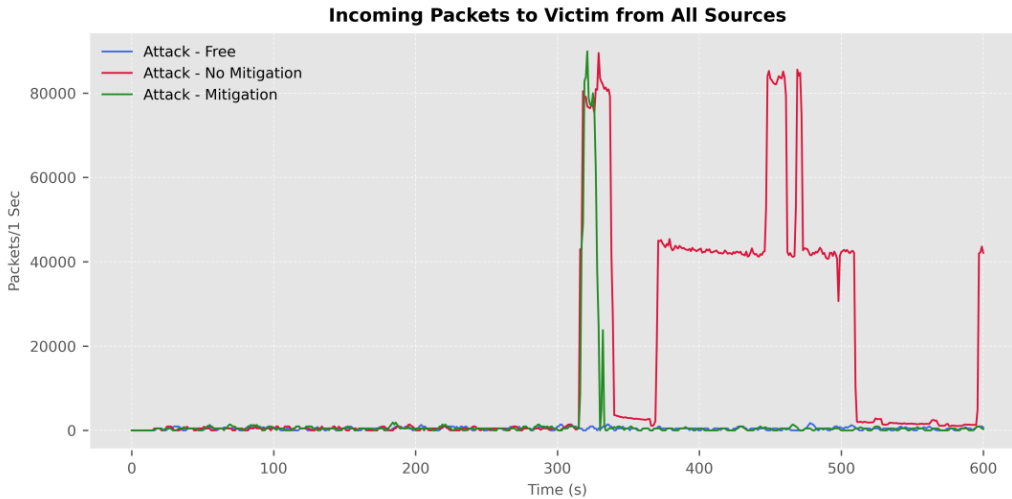

Figure 14: Incoming packets to the victim from all sources in the UDP flood attack scenario

#### 4.5.3 ICMP Flood Attack Scenario

In the ICMP flood scenario, a set of compromised hosts (h1, h9, and h13) from multiple domains launched a distributed attack targeting the h5 HTTP server with a continuous stream of ICMP echo requests. The attack intensity was configured at 40,000 packets per second, which is sufficient to saturate both the server and the control plane if left unmitigated. As illustrated in Figure 15, the attack resulted in a sharp rise in Packet-in events at the local controllers. Without the defense mechanism, the Packet-in rate continued to grow rapidly, consuming controller resources and degrading responsiveness.

When the proposed framework was enabled, the anomaly was detected at the source ports within seconds, and drop rules were enforced, which immediately suppressed the malicious traffic and restored the Packet-in rate to its baseline.

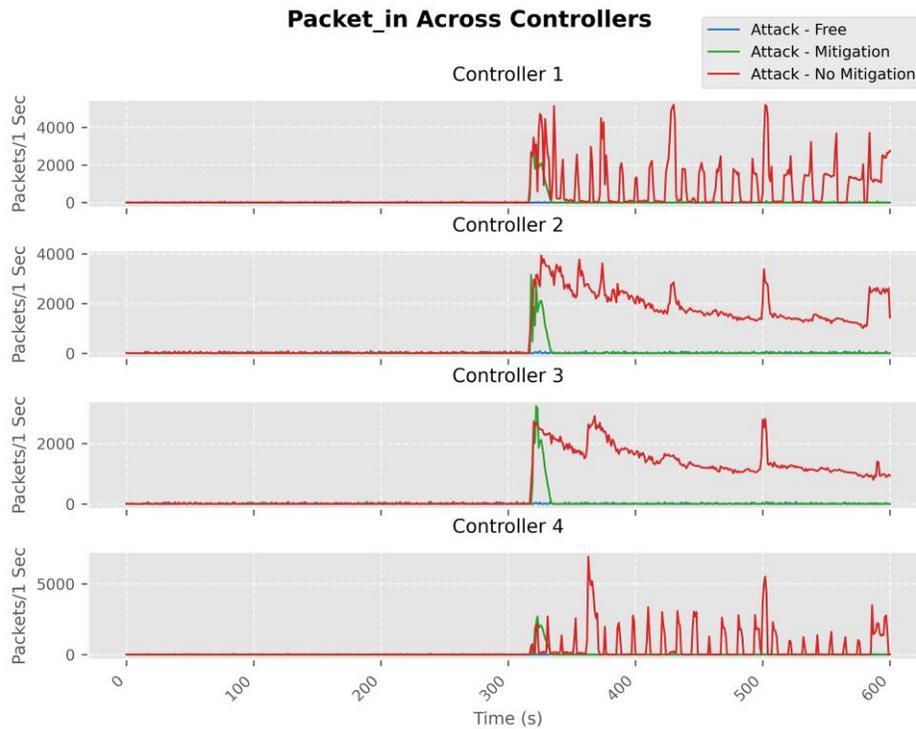

Figure 15: Packet_in across controllers in the ICMP flood attack scenario

Figure 16 presents the CPU utilization trends during the ICMP flood. In the unmitigated case, CPU usage escalated quickly due to the overwhelming number of control messages triggered by the attack. With mitigation active, CPU consumption was stabilized, remaining close to normal operating levels and ensuring continuous operation of the controllers.

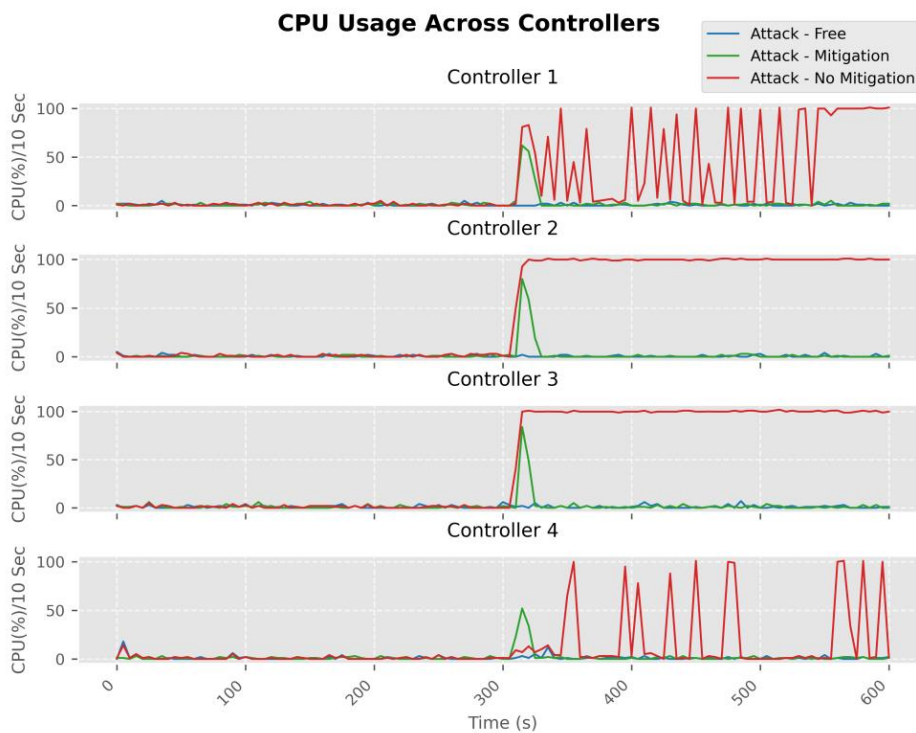

Figure 16: CPU usage across controllers in the ICMP flood attack scenario

Unlike expectations of severe synchronization disruption, the ICMP flood attack exhibited minimal impact on inter-controller synchronization, as illustrated in Figure 17. Throughout the attack, synchronization delays remained relatively stable, suggesting that the controllers were able to maintain a consistent global view of the network. This can be explained by the nature of ICMP floods, which primarily overwhelmed the victim server and generated control-plane overhead, but did not significantly interfere with inter-controller exchanges. These findings confirm that while ICMP floods threaten availability at the server and controller level, their effect on synchronization processes is less pronounced compared to other attack types.

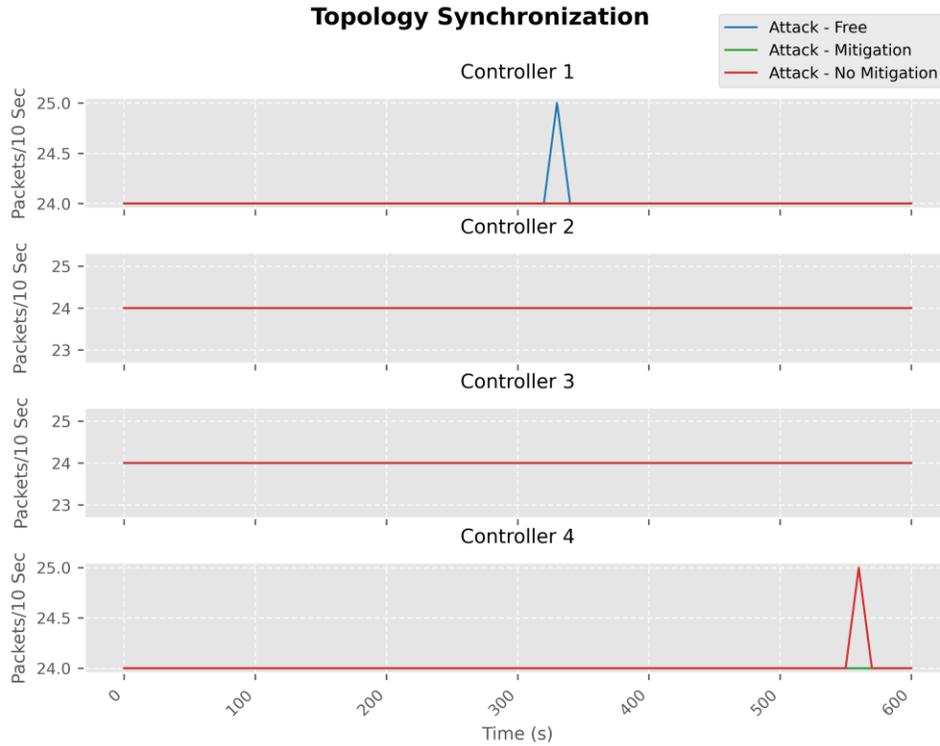

Figure 17: Peer topology synchronization in the ICMP flood attack scenario

Finally, Figure 18 illustrates the volume of ICMP traffic received by the victim server (h5). During the attack without defense, the server was inundated with ICMP requests, effectively rendering it unavailable for legitimate users. Following the activation of the defense mechanism, the incoming traffic was reduced to normal levels, ensuring uninterrupted service availability. These results confirm that the proposed framework is highly effective in mitigating ICMP flood attacks at their source, thereby maintaining stability and reliability in dSDN environments.

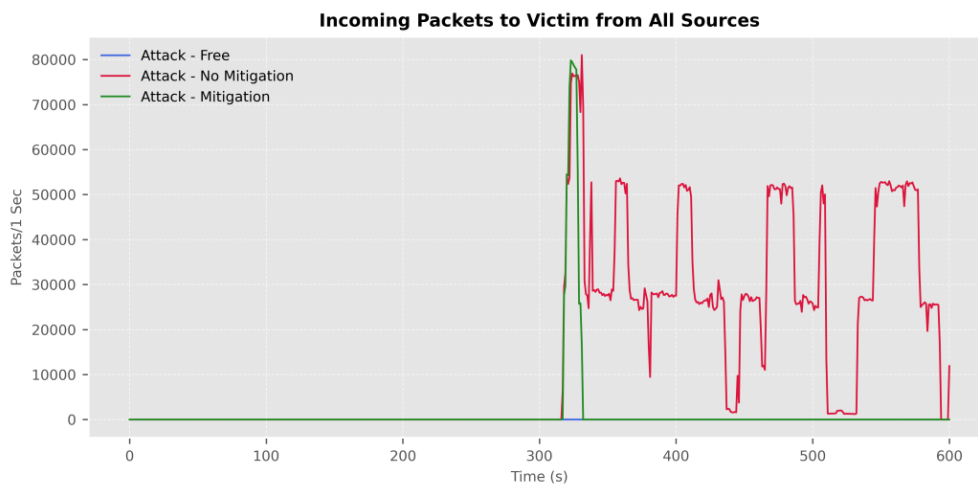

Figure 18: Incoming packets to the victim from all sources in the ICMP flood attack scenario

## 4.6 Comparison with Other Large Language Models (LLMs)

To validate the effectiveness of the proposed DeepSeek-v3 and position it within the broader landscape of LLM-based detection approaches, we compared its performance against several state-of-the-art models. These include Gemma3 [32], a lightweight and efficiency-oriented model designed for constrained environments; Qwen2 [33] and its predecessor Qwen [34], both open-source Chinese–English bilingual LLMs optimized for reasoning and general-purpose tasks; Mistral [35], a high-performance model known for its compact architecture and strong inference efficiency; LLaMA [36], a widely used family of open models released by Meta, offering strong baselines across diverse NLP tasks; and Phi4 [37], a knowledge-rich small model developed by Microsoft with emphasis on reasoning over compact size. All models were evaluated on the same test dataset of 25,933 prompts generated during the dSDN experiments, using identical evaluation metrics (accuracy, precision, recall, F1-score, confusion matrix, and ROC curves). This comparative analysis not only highlights the superior performance of DeepSeek-v3 but also reveals the strengths and limitations of alternative LLMs when applied to the task of zero-training DDoS detection in dSDN environments.

For comparative evaluation, all baseline LLMs (Gemma3, Qwen2, Qwen, Mistral, LLaMA, and Phi4) were executed through the Ollama framework on a separate evaluation server ubuntu 22.04, Linux Kernel 6.5.0-44-generic equipped with a AMD EPYC 9654, 96 cores (192 threads), 1.50–3.70 GHz CPU, 503 GiB RAM, and an NVIDIA A100 80GB PCIe, Driver 550.90.07, CUDA 12.4 GPU. This setup ensures that the reported results reflect only the inference capabilities of the models under consistent hardware conditions, independent of the environment used for the dSDN system implementation.

The comparative results, summarized in Table 4 and illustrated in the confusion matrix and ROC plots (Figures 19 and 20), highlight the clear superiority of the proposed DeepSeek-v3 model. DeepSeek-v3 achieved near-perfect performance with an accuracy of 99.99%, precision of 99.97%, recall of 100%, and an F1-score of 99.98%, while maintaining an AUC of 1.00 and misclassifying only two benign prompts. In comparison, Gemma3 also delivered strong results with high accuracy (99.60%) and an AUC of 1.00, but introduced more false positives, as seen in its confusion matrix. Qwen2 and Qwen demonstrated competitive recall but lower precision (97.71% and 96.34%, respectively), reflecting higher false-alarm rates. Mistral exhibited a noticeable performance drop with a recall of 93.03% and accuracy of 98.27%, misclassifying a significant portion of attack traffic. LLaMA reached perfect recall but struggled with precision (79.00%), resulting in a relatively high number of benign traffic incorrectly flagged as attacks. Finally, Phi4 showed the weakest performance, with an accuracy of only 63.23% and substantial misclassification, indicating limited suitability for zero-training DDoS detection. Overall, these results confirm that DeepSeek-v3 not only surpasses all baselines across every evaluation metric but also achieves a uniquely balanced trade-off between precision and recall, setting a new benchmark for LLM-based detection in dSDN environments.

Table 4: Performance comparison of DeepSeek-v3 and other LLMs using accuracy, precision, recall, and F1-score.

| Model | Accuracy | Precision | Recall | F1-score |
| --- | --- | --- | --- | --- |
| DeepSeek-v3 | 99.99% | 99.97% | 100.00% | 99.98% |
| Gemma3 | 99.60% | 98.42% | 100.00% | 99.21% |
| Qwen2 | 99.42% | 97.71% | 100.00% | 98.84% |
| Qwen | 99.06% | 96.34% | 100.00% | 98.14% |
| Mistral | 98.27% | 99.98% | 93.03% | 96.38% |
| LLaMA | 93.41% | 79.01% | 100.00% | 88.27% |
| Phi4 | 63.23% | 40.27% | 100.00% | 57.42% |

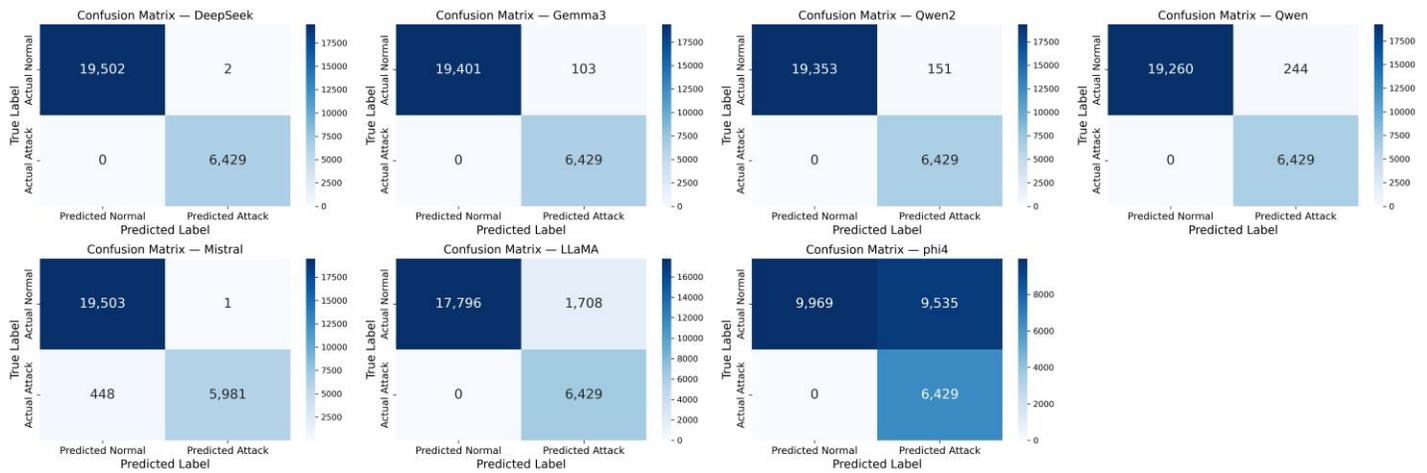

Figure 19: Confusion matrices of DeepSeek-v3 and baseline LLMs.

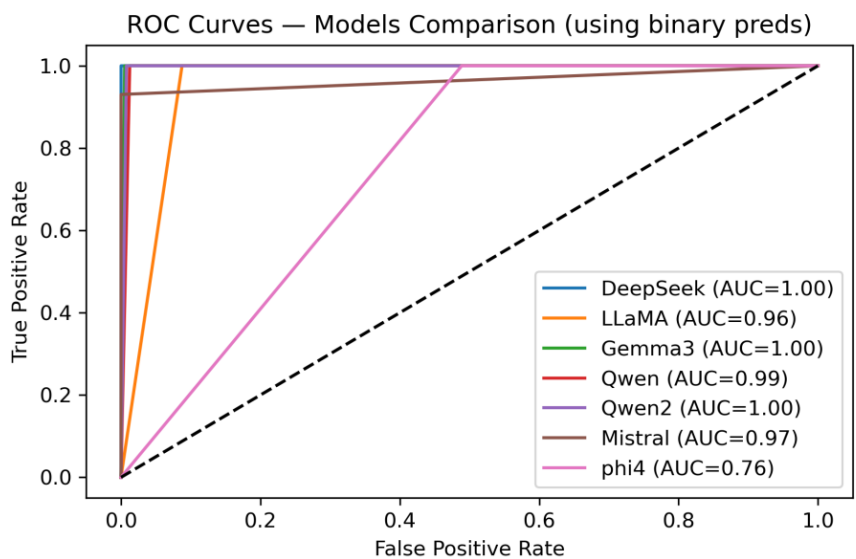

Figure 20: ROC curves of DeepSeek-v3 and baseline LLMs on the detection task.

In summary, while several state-of-the-art LLMs such as Gemma3 and Qwen2 demonstrated competitive performance, none matched the consistent reliability of DeepSeek-v3. By surpassing all baselines across every evaluation metric, DeepSeek-v3 establishes itself as the most effective zero-training LLM for proactive and scalable DDoS detection in dSDN environments. Its ability to maintain perfect recall while simultaneously minimizing false positives ensures both sensitivity and precision, qualities that are critical for preventing false alarms, safeguarding legitimate traffic, and preserving the local controllers' stability.

**4.7 Comparison with Existing Studies**

A comparative analysis with prior works reveals fundamental differences between existing approaches and our proposed framework. As summarized in Table 5, nearly all previous studies focused exclusively on cSDN environments and relied heavily on complex flow-level or hybrid feature sets, combined with machine learning or deep learning models that require extensive training. While these methods achieved competitive performance, they lacked support for early and source-based detection and often introduced high complexity. In contrast, our study is among the first to target dSDN and departs from the training-intensive paradigm by employing a zero-training LLM approach with lightweight port-level features. This design shift not only reduces system complexity but also enables proactive and distributed detection that prior works did not address.

Table 5: Comparison of our proposed framework with the existing literature

| Study | SDN Type | Data Type | Training | Detect attack from source | Early detect | Complexity | Accuracy | Precision | Recall | F1-score |
|---|---|---|---|---|---|---|---|---|---|---|
| [19] | cSDN | Flow-level | Yes | No | No | High | 99.99% | 99.99% | 99.99% | 99.99% |
| [21] | cSDN | Flow-level | Yes | No | No | High | 99.98% | 99.97% | 99.98% | 99.97% |
| [22] | cSDN | Flow-level + Header-level | Yes | No | No | High | 99.91% | 99.89% | 99.91% | 99.89% |
| [20] | cSDN | Flow-level | Yes | No | No | High | 99.92% | 99.90% | 99.90% | 99.90% |
| [24] | cSDN | Flow-level | Yes | No | No | High | 99.38% | 99.41% | 99.40% | 99.39% |
| [26] | cSDN | Flow-level + Port-level | Yes | No | Yes | High | 98.80% | 98.27% | 97.91% | 97.65% |
| [27] | cSDN | Flow-level | Yes | No | No | High | 98.30% | 97.72% | 97.73% | 97.70% |
| [28] | cSDN | Flow-level | Yes | No | No | High | 97.00% | NA | 96.00% | NA |
| [29] | cSDN | Flow-level | Yes | No | No | High | 95.24% | NA | NA | NA |
| Our Study | dSDN | Port-level | No | Yes | Yes | Low | 99.99% | 99.97% | 100% | 99.98% |

For instance, [24, 27] proposed flow-level detection methods in cSDN, achieving reasonable accuracy but with high computational complexity and no support for early or source-based detection. Similarly, [28, 29] adopted flow-level statistics with supervised learning, but their approaches suffered from scalability limitations and delayed response due to the reliance on fully established flows. [26] combined flow-level and port-level features, offering partial improvements, yet their model still required offline training and did not support direct mitigation at the source. More recent works, such as [19, 20], reported strong detection results in cSDN, but they continued to depend on complex models and centralized infrastructures. Compared with these studies, our framework uniquely integrates lightweight port-level monitoring with zero-training LLM inference in a dSDN, enabling early, source-based detection with significantly lower complexity while maintaining state-of-the-art accuracy.

Overall, the comparative analysis demonstrates that our proposed framework achieves results that are either comparable to or superior to existing state-of-the-art studies, despite its lower complexity. Unlike prior works that rely on flow-level or hybrid features, centralized architectures, and extensive model training, our approach introduces a lightweight, training-free solution tailored for dSDN. By combining port-level monitoring with zero-training LLM inference, the system not only attains near-perfect accuracy (99.99%) and recall (100%) but also provides early and source-based detection capabilities that were absent in previous studies. These advantages highlight the novelty and practical relevance of our contribution, establishing it as a promising direction for scalable and proactive DDoS defense in dSDN environments.

## 5. Conclusion and Future Work

This study presented a novel framework for detecting and mitigating DDoS attacks in dSDN environments by leveraging port-level feature extraction, prompt engineering, and the zero-training capabilities of the DeepSeek-v3 LLM. Unlike conventional approaches that rely on flow-level features and retraining, our design capitalizes on lightweight port-level statistics aggregated every ten seconds and encoded into natural language prompts enriched with benign examples for in-context learning. This integration enables early, source-based detection directly at the attacker's port, followed by immediate mitigation through automated controller actions. Experimental evaluation across three flood-based attack scenarios (TCP, UDP, and ICMP) demonstrated that the proposed system achieves near-perfect detection performance, with an accuracy of 99.99%, recall of 100%, and an exceptionally low false-alarm rate. Furthermore, the distributed design of the framework ensures low-latency response and scalability across multiple domains. Collectively, these results validate the effectiveness of combining LLM-driven inference with decentralized control, establishing a robust and proactive defense mechanism for modern SDN infrastructures.

For future work, several promising directions can be pursued to extend the proposed framework. First, we plan to evaluate the system on larger and more complex network topologies in order to assess its scalability and performance under higher traffic volumes and more diverse inter-domain interactions. Second, while this study focused primarily on flood-based DDoS scenarios, future experiments will investigate the detection and mitigation of additional attack types. Third, an important step will be testing the framework in real-world or production-like environments, where heterogeneous devices, dynamic traffic conditions, and multi-tenant services can provide a more realistic validation of its practicality. By addressing these directions, the proposed system can be further strengthened and generalized to meet the demands of modern and evolving network security landscapes.

**Funding:** This work was partially supported by the National Natural Science Foundation of China (Grant No. 62171291).